\documentclass[12pt]{article}
\usepackage{graphicx}
 \usepackage{longtable}
\relax
\usepackage{amssymb}

\newcommand{\bo}{{\bar o}}

\def\bo{{\raise.15ex\hbox{\large$\Box$}}}               % D'Alembertian

\def\face{{\raise.2ex\hbox{$\displaystyle \bigodot$}\mskip-2.2mu \llap {$\ddot
        \smile$}}}                                      % happy face
\def\Zbf{{\bf Z}}
\def\leftrightarrowfill{$\mathsurround=0pt \mathord\leftarrow \mkern-6mu
        \cleaders\hbox{$\mkern-2mu \mathord- \mkern-2mu$}\hfill
        \mkern-6mu \mathord\rightarrow$}       % <--> double differential
\def\dvec#1{\vbox{\ialign{##\crcr
        \leftrightarrowfill\crcr\noalign{\kern-1pt\nointerlineskip}
        $\hfil\displaystyle{#1}\hfil$\crcr}}}           % <--> accent
\def\beq{\begin{equation}}
\def\eeq{\end{equation}}

\def\beqx{\begin{displaymath}}
\def\eeqx{\end{displaymath}}

\def\beql{\begin{eqnarray}}
\def\eeql{\end{eqnarray}}

\newcommand{\Tr}{{\rm Tr}}

\newcommand{\bea}{\begin{eqnarray}}
\newcommand{\eea}{\end{eqnarray}}

\newcommand{\mod}{\;{\rm mod }\;}

\def\[{\left [}
\def\]{\right ]}
\def\({\left (}
\def\){\right )}

\def\+{\oplus}

\begin{document}

%\vspace*{0.15in}
\hbox{\hskip 12cm NIKHEF/2009-021  \hfil}
\hbox{\hskip 12cm IFF-FM-2009/03  \hfil}
%\hbox{\hskip 12cm hep-th/yymmnnn \hfil}

\vskip .5in

\begin{center}
{\Large \bf Heterotic Weight Lifting }

\vspace*{.4in}
{B. Gato-Rivera}$^{a,b}$\footnote{Also known as B. Gato}
{and A.N. Schellekens}$^{a,b,c}$% \footnote{t58@nikhef.nl} 
\\
\vskip .2in

${ }^a$ {\em NIKHEF Theory Group, Kruislaan 409, \\
1098 SJ Amsterdam, The Netherlands} \\

\vskip .2in

${ }^b$ {\em Instituto de F\'\i sica Fundamental, CSIC, \\
Serrano 123, Madrid 28006, Spain} \\

\vskip .2in

${ }^c$ {\em IMAPP, Radboud Universiteit,  Nijmegen}

\end{center}

\begin{center}
\vspace*{0.3in}
{\bf Abstract}
\end{center}

We describe a method for constructing genuinely asymmetric (2,0) heterotic strings out
of N=2 minimal models in the fermionic sector, whereas the bosonic sector is only partly build
out of N=2 minimal models. This is achieved by replacing one minimal model plus the superfluous $E_8$ factor by
a non-supersymmetric CFT with identical modular properties. This CFT generically lifts the weights in the 
bosonic sector, giving rise to a spectrum with fewer massless states. We identify more than 30 such lifts, and we expect many more to exist. This yields more than 450 different combinations.
Remarkably, despite the lifting of all
Ramond states, it is still possible to get chiral spectra. Even more surprisingly, these chiral spectra include
examples with a certain number of chiral families of $SO(10)$, $SU(5)$ or other subgroups, including just $SU(3) \times SU(2) \times U(1)$.
The number of families and mirror families is typically smaller than in standard Gepner models. Furthermore, in a large number
of different cases, spectra with three chiral families can be obtained.  
Based on a first scan of about $10\%$ of the lifted Gepner models we can construct, we
have collected more than 10.000 distinct spectra with three families, including examples without mirror fermions.
We present an example where the GUT group is completely broken to the standard model, but the
resulting and inevitable fractionally charged particles are confined by an additional gauge group factor.

\vskip .5truecm

\noindent
October 2009

\newpage

During the last part of the eighties of last century, several approaches were developed  to construct 
genuine heterotic strings using exact conformal field theories. By ``genuine" we mean that the bosonic
sector contains no superfluous fermionic remnants, such as an $E_8$ ``hidden" gauge group, an extension of
the standard model gauge group to $E_6$ or an N=2 worldsheet symmetry without any function. 
Such vestiges of fermionic symmetries are often present because of the difficulties imposed by modular
invariance. Ideally, one would like to build the fermionic sector out of N=2 building blocks, and the bosonic
sector out of N=0 building blocks. For arbitrary choices of building blocks, the (extended) Virasoro characters have totally
different modular transformations, and there is no way to write down a modular invariant combination. Essentially
two ways of getting around this are known. One way is to first build a symmetric type-II string theory out of N=2 building blocks, 
and then to map one fermionic sector to a bosonic one using the bosonic string map \cite{LLS}. This uses the observation
that the covariant NSR characters together with the superghosts transform in exactly the same way under
all modular transformations as the characters of $SO(10)\times E_8$. 
This includes as a special case the earlier idea of ``embedding the holonomy group in $E_8$", used in 
Calabi-Yau compactifications \cite{Candelas:1985en} and the first orbifold compactifications, and was applied by Gepner to his CFT
construction \cite{Gepner:1987qi}. The other way around the obstacle of modular invariance is to use very simple building blocks
that can be tailored to meet the requirements of the fermionic sector. Those building blocks are then typically limited
to combinations of free bosons and free fermions, with a variety of boundary conditions; early examples are in
\cite{Narain:1985jj},  \cite{Kawai:1986va}, \cite{LLS}, \cite{Antoniadis:1986rn} and \cite{Narain:1986qm}.

Obviously both approaches have their intrinsic limitations. The former, also known as (2,2) compactifications,
miss a huge part of the full set of (2,0) heterotic strings. The latter are limited to very special conformal field theories,
which cannot reasonably be expected to be representative for the full heterotic landscape. In both cases there is
a risk about being misled by artefacts due to the limitations of the particular construction. Although that risk is hard to
overcome anyway, it is important to try and push the boundaries further and observe which features remain, and which ones
disappear.

Gepner's original construction had the virtue of overcoming the limitation to free CFT's of earlier exact CFT constructions,
but was still a (2,2) construction. In \cite{Schellekens:1989wx} an attempt was made to partly overcome that
restriction, by keeping the same $N=2$ minimal model building blocks, but relaxing the constraints of space-time
and world-sheet supersymmetry in the bosonic sector. This was possible because simple current modular
invariants\footnote{Simple current symmetries in conformal field theory were discovered in 
\cite{Schellekens:1989am} and \cite{Intriligator:1989zw}.
Here we will make use of the formalism developed in \cite{GatoRivera:1991ru} and
\cite{Kreuzer:1993tf}, which gives the complete set of simple current MIPFs (Modular Invariant Partition Functions), and which was not yet available in 1989.}  are in general not symmetric. World-sheet and space-time supersymmetry appear as chiral algebra extensions
in the fermionic sector, but these can be mapped to higher spin extensions in the bosonic sector. This avoids the extension
of $SO(10)$ to $E_6$, and vastly enlarges the number of possibilities. It was also pointed out in \cite{Schellekens:1989wx}
that one could go a step further: break the $SO(10)$ group into subgroups (such as $SU(3)\times SU(2)\times U(1)^2$), and
only reassemble it in the fermionic sector, where it is needed for the bosonic string map. This idea was not
pursued in 1989, because it required huge amounts of computer power not available at that time. It led to a general
proof of the inevitability of fractional\footnote{Throughout this paper, ``fractional charge" refers to charges of elementary or
QCD-composite color singlets, ignoring all interactions beyond the standard model. In other words, to representations of the Lie algebra $SU(3)\times SU(2) \times U(1)$ that are not representations of the group $S(U(3)\times U(2))$.}
 charges  \cite{Schellekens:1989qb}
(not necessarily light, and perhaps confined by additional interactions)
in such models. The idea was taken up again in 1995 by Blumenhagen et al. 
\cite{Blumenhagen:1995tt}\cite{Blumenhagen:1995ew}\cite{Blumenhagen:1996xb}\cite{Blumenhagen:1996gz}\cite{Blumenhagen:1996vu}. Around the same time several other papers on $(2,0)$ models appeared
({\it e.g.}   \cite{Kachru:1995em} \cite{Distler:1995bc} \cite{Berglund:1995dv}) with a less obvious relation to
the ideas discussed here. There was also a recent discussion in
 \cite{Kreuzer:2009ha}. In string phenomenology heterotic  strings have undergone a revival recently (see {\it{e.g.}} \cite{Lebedev:2006kn} \cite{Faraggi:2006bc}), but these works are based on free CFT's .

Recently we have returned to the idea put forward in \cite{Schellekens:1989wx},
to see if with current technology we could answer some questions
that were left pending in 1989, such as the nature of the fractional charges, the distribution of families, and
the family number quantization encountered in previous work.  
The net number of families in the aforementioned
simple current modified
Gepner models, still with unbroken $SO(10)$, was found to be always a multiple of 4, 6, or larger even numbers. The
only known way around this was found by Gepner \cite{Gepner:1987hi}, and requires the use of three exceptional (E-type) modular invariants
of $SU(2)$ level 16, in the tensor product (1,16,16,16). This case was also studied in \cite{Schellekens:1989wx}, and
led to a list of about 40 3-family models with a gauge group $SO(10)$ and $E_6$. In a forthcoming paper \cite{GS} we
will present the results of a new and much more detailed analysis. In particular this includes a negative answer to  a question raised in the conclusions
of \cite{Schellekens:1989wx}: will the breaking of $SO(10)$ to smaller subgroups (the
standard model or one of its extensions) affect the family quantization? We find
that it does, but that it remains even (and usually a multiple of six), and hence still rules out three families for all
simple current modular invariants. We will also present an analysis of the exceptional three-family case 
under the same conditions. The list of 40 models of  \cite{Schellekens:1989wx} is enlarged to almost 1000.

The purpose of the present paper is to try something entirely different, namely to modify the N=2 building blocks 
themselves. The idea is to replace an N=2 building block by a different one with identical modular properties. This
replacement can then be made in the bosonic sector, leaving the fermionic one intact, and respecting modular invariance.
Hence all CFT conditions for sensible four-dimensional heterotic strings are respected.  

Finding two distinct CFT's with identical modular properties is not easy. There are plenty of examples 
for {\it free} theories, such as the WZW models $D_n$, level 1, for values of $n$ differing mod 8; $E_8 \times E_8$ and
$SO(32)/\Zbf_2$ used in the heterotic string in ten dimensions, and the aforementioned bosonic string map. There are
also examples for {\it interacting} CFT's, namely the meromorphic conformal field theories classified in \cite{Schellekens:1992db}, but in that case
the resulting CFT is trivial. Here we are looking for examples of interacting CFT's with large numbers of characters.

If one tries to replace an N=2 minimal model by a CFT with the same central charge the number of possibilities
is very limited, although an example may well exist, because without specific extended symmetries most
models lie beyond the scope of the minimal Virasoro models. 
Instead of trying that,
we make use of the fact that in the bosonic sector there is still a superfluous $E_8$ factor which we can put to good use.
So instead of replacing an $N=2$ building block with central charge $c$, we replace it by a building block with central
charge $c+8$ and remove the $E_8$ factor. This should not be confused with other uses of the
$E_8$ factor. In particular, along the lines of \cite{Schellekens:1989wx} one may deconstruct the
$E_8$ into smaller pieces and use asymmetric simple current MIPFs to reassemble it in the
fermionic sector, where it is needed. One could also consider embedding (part of) the standard gauge
group in the $E_8$ factor. Some of this has already been tried, for example in \cite{Blumenhagen:1996xb}, but this is not what we will be doing in this paper. This does not mean that our results are
totally unrelated to those of other methods. The history of string theory contains plenty of warnings
against that sort of statement.

Examples of what we are looking for
can be found as follows (without any claim of generality). A minimal\footnote{The main idea can be generalized straightforwardly to the non-minimal  $N=2$ models described by Kazama and Suzuki in  \cite{Kazama:1988qp}.} $N=2$ model can be
obtained using the coset construction as
$$   {SU(2)_k \times U_4  \over U_{2(k+2)}} \ . $$
Here $U_N$ denotes the compactified free boson with a radius such that the total number of primary fields is $N$ 
(which is always even). The second factor, $U_4$, is often denoted as $SO(2)$. This coset is subject to field
identification by the simple current $(J,2,k+2)$, where $J$ is the $SU(2)$ simple current, and the $U_{N}$ fields
are labelled by their charges as $0,\ldots,N-1$, where ``0" is the vacuum.  To describe the coset
field identification as a simple current extension we are formally treating the denominator factor as a CFT with
complex conjugate $S$ and $T$ matrix. A CFT with complex conjugate $S$ and $T$ matrices is sometimes called  the
``complement", because it can be combined with the original to produce a meromorphic CFT with just one primary field,
whose central charge is a multiple of eight.
The field identification current of a coset CFT has spin 0, and may be thought of formally as an extension of the vacuum module.

To manufacture a new CFT with the same modular transformation properties we remove first the field identification
extension, then we tensor with $E_8$, then we mod out $E_8$ by $U_{2(k+2)}$, and then we restore the
effects of the field identification as a standard extension of the resulting CFT. This can be done provided we can find
a suitable embedding of the $U_{2(k+2)}$ factor in $E_8$. We can think of several other ideas that might
work similarly, such as extending $SO(2)$ to $SO(18)$ and embedding the $U(1)$ in it, but we will
focus here on what appears to be the simplest possibility. We will also focus on cases where the $E_8/U(1)$ coset is itself
an already known CFT, so that there is no need to compute non-trivial branching functions.

It is simplest to illustrate this with an example, which is in fact the first one we considered. Since the standard model
can be embedded in $E_8$ (via $E_6$), the standard model $Y$ charge is an example of a $U(1)$ factor that
can be embedded in $E_8$. The embedding is
$$E_8 \supset  A_{2,1} A_{1,1} A_{4,1} U_{30} \ ,
 $$
where $A_{n,m}$ denotes $A_n$ at level $m$. The first two factors are simply the $SU(3)$ and $SU(2)$ of the standard
model, the $A_4$ factor comes from combining $SO(6)$ (the commutant in $E_8$ of $SO(10)$)  with $B-L$. Finally, $U_{30}$
is the standard model $Y$ charge, normalized in the standard GUT way. 

We  now expect that
\beq {E_8 \over U_{30}} =   A_{2,1} A_{1,1} A_{4,1} \ , \eeq
for a suitable primary field assignment.
It is straightforward to check that this is indeed true. Note that the modular transformation matrices of the CFT's
on both sides of the equation are just phases, and it is obvious that a mapping of the fields exists so that they coincide. The
same would also be true for, say, $A_{1,1}A_{10,1}$ in comparison to the complement of $U_{22}$. However, it will in general
not be true that the T-matrices can be matched. The embedding in $E_8$ ensures that.

Having identified the complement of $U_{30}$, we can now assemble the new CFT. The solution to $2(k+2)=30$ is $k=13$,
and hence we expect to be able to construct a new CFT that transforms as the minimal model with $k=13$. So consider
\beq \label{lift13} A_{1,13} U_4 A_{2,1} A_{1,1} A_{4,1} \eeq
We still have to re-introduce the equivalent of the field identification current. For this purpose we 
extend the chiral algebra with the order-2 current $(J,2,0,J,0)$. Now, as a check, we compare the resulting matrices
$S$ and $T$ with those of the $N=2$ minimal model with $k=13$, after determining the correct mapping of the primaries,
and we find that they are indeed identical. We will refer to the new CFT described here as the ``lift" of the minimal $N=2$, $k=13$ model.

The conformal weights of a coset CFT $G/H$ are equal to $h^G_i - h^H_j$ modulo integers. These integers are zero if
the ground state Lie algebra representation of $i$ contains the ground state Lie algebra representation of $j$, and otherwise they are positive. If we replace
the denominator factor $H$ by a numerator factor $H^c$ (where $c$ stands for ``complement"), to obtain a tensor CFT
$G\otimes H^c$, then the weights of the new CFT are $h^G_i + h^{H^c}_j$, with $h^{H^c}_j = - h^H_j \mod 1$. Note however 
that both $h_j^{H}$ and $h^{H^c}_j$ are positive, so that the weights of the lift are larger than the
weights of the original,  $h^G_i + h^{H^c}_j > h^G_i - h^H_j $ in all cases where the ground state $j$ is contained in  $i$. 
It is for this reason that we use the term ``lift".
In all other cases the difference may go either way, and furthermore field identification complicates the discussion. In the coset
CFT, the field identification current relates fields of equal weight, where in the lift it acts like an extension that combines representations of
(in general) different weight.

The potential advantages of weight lifting should be obvious. All exact string theory constructions suffer from a plethora of
superfluous states. In heterotic strings, the number of  families tends to be too large, usually there are mirror fermions, there
are large numbers of moduli and other gauge singlets, and if grand unification is broken there will be fractional charges.
Furthermore, if we break supersymmetry we are in general faced with tachyons. One cannot reasonably expect all these problems
to be solved in one step, but there is at least a chance of moving in the right direction.

So let us see if the lift lives up to its name in a concrete example.
The minimal $N=2, k=13$ CFT has 420 primary fields. If we replace it by its lift, the following happens to the
conformal weights of the field.  Here we  will only consider primaries with $h \leq1$ either before and/or after the lift,  {\it i.e.} those primaries that can
contribute to the massless sector. Reshufflings of the massive states will also occur, but are less interesting. 
We find that the conformal weight of 136 primaries is lifted above $h=1$, whereas the conformal weight of  81 is lowered below $h=1$. Furthermore there are 37 primaries with $h \leq1$ before and after the lift.

Taking into account the effects of a lift in the spectrum is straightforward. Given a complete spectrum of a Gepner combination
with some choice of modular invariant partition function, one simply replaces all 
weights in the bosonic sector of the heterotic string by the weights of the
lift. Obviously this requires knowledge of the full partition function including massive states in the bosonic sector, but it may be
restricted to the massless states in the fermonic sector. Furthermore it is -- obviously --  crucial to take into account the correct ground state
dimensions of all the primaries of the lift.

Among the primary fields that are lifted are all Ramond states. At first sight this might appear to be disastrous. Consider for example
the combination $(4,4,8,13)$.
The 
diagonal MIPF of this Gepner model gives rise to a spectrum with 75 chiral fermions in the representation (27) of $E_6$,
3 in the $(\bar{27})$ and 450 singlets. If we lift the $k=13$ factor (the resulting tensor CFT will be denoted as $(4,4,8,\widehat{13})$),
the spinor weight that extends $SO(10)$ to $E_6$ is lifted, and hence we get an $SO(10)$ gauge group instead of $E_6$. That
is a good feature in itself, but it also implies that the simplest realization (the diagonal MIPF) of these models is not directly
linked to any Calabi-Yau-type compactification.
However, unfortunately all chiral matter is lifted as well, and the surviving massless matter consists of 20
vectors and 1088 singlets of $SO(10)$. The multiplicity 20 is a combination of 12 singlets of the lift  gauge group, eqn. (\ref{lift13}), plus one
eight-dimensional representation consisting of the spin-$\frac32$ representation of $A_{1,13}$ combined with the spin-$\frac12$ 
representation of $A_{1,1}$. The 1088 singlets are an assembly of many non-trivial representations of the lift CFT. Note that despite
the weight lifting, the total number of massless representations does not really decrease. This is because the ground state
dimensions of the unlifted or lowered primaries usually increases, because they are in various
non-trivial representations of semi-simple Lie algebras.

This example may suggest that this disappearance of chiral matter is a general phenomenon. However, one may consider a much
more general set of MIPFs than just the diagonal one, as was done in \cite{Schellekens:1989wx}. As mentioned above, in  \cite{GS} we will
take that one step further, and consider all (2,0) Gepner models (unlifted) with $SO(10)$ broken to $SU(3)\times SU(2) \times U(1)$
and without world-sheet supersymmetry in the bosonic sector. This opens up the possiblility of pairing spinors of $SO(10)$
(or their subgroup decomposition) with NS states rather than Ramond states, and evade the lower bound $h \geq \frac{c}{24}$ of
the latter. Since the entire formalism is already available and well-tested on the standard Gepner models, we
could straightforwardly apply it to the lifted Gepner models. With all these restrictions removed, the total number of 
MIPFs is enormous. We are currently not able 
to scan them all exhaustively, because there are too many possibilities. All statements that follow are based on random samples.

In this paper we will only consider cases where supersymmetry is not broken, and where the standard model
is embedded in the canonical way in $SO(10)$. This means that gauge couplings will satisfy the usual Susy-GUT relations, including
the usual slightly problematic gap between the string scale and the GUT scale. Allowing broken supersymmetry would be straightforward, but one has to get rid of the tachyons. We may consider that possibility in the future. 

When we started using the $k=13$ lifted CFT in a few examples with non-diagonal MIPFs, we immediately obtained astonishing results.  
The first example we tried, the $(4,4,8,\widehat{13})$ used above, {\it did} give rise to chiral matter with
these more general MIPFs. We found examples with 2, 4, 6, 8, 12, 14 and 28 net chiral families of $(16)$'s of $SO(10)$ or subgroups. 
So chiral matter {\it does} appear, but the multiplicities are still only even.
 Then
we tried $(1,3,3,3,\widehat{13})$. In this case the number of families turns out be quantized in units of 1, and in particular we
encountered cases with $1,\ldots,12, 14, 24$ and 36 families. Hence here, for the first time, we get 3 families from N=2 minimal models without
using exceptional MIPFs! The simplest one we encountered has gauge group $SU(5)$ with a spectrum $3 (10) + 12 (\bar 5) + 
9 (5)$ plus 575 singlets.  Note the absence of mirrors
in the $(\bar{10})$, another feature that up to now was hard to obtain in standard Gepner models.
In addition to $SU(5)$ there is of course a large gauge group from the lift CFT and the  minimal 
model $U(1)$'s. 
All particles may be in non-trivial representations of that group;
we will give some more details in a different example below. The next lifted Gepner model we considered was $(3,4,13,\widehat{13})$.  Again we found the
family number quantized in terms of integers, this time with values $1,\ldots,8, 10, 11,13,18,$ and  30. The simplest three family model we found is $3 (10) + 9 (\bar 5) + 6 (5) + 395 (1)$, again with gauge group
$SU(5) \times G_{\rm extra}$. 
There are 16 Gepner models containing a $k=13$ minimal factor, but we do not intend to present a full enumeration in this paper. 

We have several checks on our computations. First of all, the lifting procedure is a very simple alteration of the computation
for normal Gepner models. One of us was involved in extensive computations of this kind in 1989  \cite{Schellekens:1989wx}, and the results of
that computation are still available and serve as a check. However the original computer code was lost and had to be
rebuild. Our new, more extensive results for the unlifted Gepner models are therefore independent of the 1989  results, and 
they are in full agreement with them, as well as with other results of that time \cite{Lutken:1988hc} \cite{Fuchs:1989yv}.
Furthermore anomaly cancellation is an important check. This entire class of models, both
lifted and unlifted, belongs to the class where the results of \cite{Schellekens:1986xh} apply. Indeed, since
we do not know how to get these spectra from any kind of compactification, one cannot derive the
anomaly cancellation from the ten-dimensional one. To be certain how the anomalies cancel in this
case, the derivation of \cite{Schellekens:1986xh}, which relates anomaly cancellation directly
to modular invariance, is essential.
Because of modular invariance, for very model the entire anomaly must factorize
into a single $\Tr F^2 - Tr R^2$ factor times a polynomial linear in $\Tr F$. The latter is unrestricted, in particular it may vanish,
and it may vanish in any subgroup of the full gauge group. But the factor $Tr F^2$ must appear for every factor of the gauge group
with exactly this normalization (assuming it is computed in the vector representation;  otherwise the appropriate correction factor
must be added). 
In the unlifted Gepner models
there is always an unbroken $E_8$ factor in the gauge group, under  which only the anomaly-free gauginos are charged.
Hence it cannot appear in $\Tr F^2$ and hence {\it all} anomalies must cancel. In the lifted Gepner models there is
no $E_8$, and hence factorizable anomalies may appear. This is indeed what happens.  Sometimes the standard model
$Y$ charge is anomalous, and then the spectrum is rejected. It may also happen that $B-L$ is anomalous. In both cases
we have checked factorization of the anomaly to make sure that modular invariance is respected. This turned out to be an
extremely useful test to detect trivial mistakes. 

An anomalous $B-L$ is a mixed blessing in heterotic strings, because on the one hand it makes the undesirable 
$B-L$ vector boson massive, but on the other hand they  generate Fayet-Illiopoulos terms that tend to drive the theory
away from the RCFT point. At the very least this is an issue that requires further attention.
In the first of the $SU(5)$ models mentioned above, $B-L$ is anomalous,
in the second it is not, but there is a nearly identical model with 469 singlets instead of 395 where $B-L$ is anomalous.

Which other
minimal models can be lifted? We stumbled on the $k=13$ example by means of the standard model $Y$ charge embedded, via $SO(10)$,
in $E_8$. We can just as well consider $B-L$ instead of $Y$. This turns out to be a $U_{20}$, which can be used in a completely
analogous way to lift the $k=8$ minimal model. Going through a list of maximal subgroups of $E_8$ rapidly produces some more examples. 

A complete enumeration of an interesting subclass can be obtained as follows. The $U_M$ factor of interest, combined with its complement, must contain
the Cartan sub-algebra of $E_8$ level 1, which consists of eight commuting free bosons.  If the free boson in $U_M$ is itself one of those
eight bosons, the complement must have seven additional ones, and since it has central charge seven it must consist entirely out of free bosons compactified on a 7-torus. The only way out of that conclusion would be to realize some of the $E_8$ Cartan sub-algebra generators 
as non-trivial vertex operators that carry non-vanishing $U_{M}$ charges. This would be analogous to what happens if one builds
$E_8$ out of the subgroup 
$G_2\times F_4$, which has total rank 6. It is not clear if such a situation can occur if one of the factors is a $U(1)$, but we will ignore that 
possibility here. Then the class of complements we consider must be a
product of simply laced algebras and $U(1)$'s. Note that this implies that the rank of the gauge group does not change: a minimal model
combined with $E_8$ gives rise to a gauge group $U(1)\times E_8$, whereas a lifted level $k$ minimal model yields
a gauge group $A_{1,k} \times U_4 \times X_7$, where $X_7$ is the gauge theory of  the seven-torus.

The complement may be an unextended tensor product or a simple current extended one. 
In the former case a full classification is easy, since the combined simple current symmetries of the tensor product must
exactly match those of the $U_N$ factor. In particular
the result must have a $Z_N$ simple current symmetry. This $Z_N$ symmetry
may be distributed over separate factors of the rank-7 theory, but each prime factor and its powers must belong to just one factor. Hence
there can be at most one factor $U_M$ in the complement, since $M$ is always even. Consider first the easier case without $U_M$
factors. It must be a tensor product of simply-laced affine Lie-algebras with mutually prime centers. Furthermore $D_{2n}$ is not allowed, 
since its center is $\Zbf_2 \times \Zbf_2$. The possibilities that remain are $E_7$,  $A_7$,  $D_7$, $D_5\times A_2$ and  $A_4 \times A_3$, all at level 1. 
Note that $A_{1,1} = U_2$; we will consider this below. The simple current group of these simply laced algebras directly determines
the $U_N$ factor for which it is a candidate complement, and it is easy to check that this works as expected. These algebras provide
complements
for $k=-1, 0, 2, 4$ and  $8$, where $k=-1$ and $0$ are only formal solutions. 
 
If there is one $U_M$ factor (but still no extension), we need a rank-6 simply-laced theory with mutually prime centers that are all odd. The possibilities
are $E_6, A_6$ and $A_4\times A_2$. Here we are faced with an {\it a priori} infinite range of values of $M$, limited only
by the requirement that $M$ is not a multiple of 3 in the first case, 7 in the second and 3 or 5 in the third. Hence the possibilities are
\begin{eqnarray*}
\hfill E_6 \!&\!  \times&\!\!U_M \times U_{3M},  \       \  M \not=0\hbox{~mod~}3 \\
A_6 \!&\! \times&\!\!U_M \times U_{7M},     \      \     M \not=0\hbox{~mod~}7 \\
A_4 \times A_2\!&\!\times &\!\!U_M \times U_{15M},  \  M \not=0\hbox{~mod~}3 \hbox{~and~} M \not=0\hbox{~mod~}5 \\
\end{eqnarray*}
The values of $M$ are restricted by the requirement that it must be possible to extend the chiral algebra to $E_8$. A necessary
and sufficient condition is the existence of an integer spin simple current of order $\ell M$, where $\ell=3,7,15$ respectively. Since the
total number of primaries prior to the extension is the square of this number, any extension by such a current reduces the number of primaries to 1, and
then the only possible solution is $E_8$. Without loss of generality we may assume that the current has charge 1 in the last factor, which 
contributes ${1\over 2 \ell M}$ to the conformal weight. This can only add up to an integer if the current has maximal order in each of
the other factors. In the first case we get then the condition
$$ \frac23 + \frac{q^2}{2M} + {1\over 6  M} \in \Zbf $$
The solutions are $M=12 p + 2, p \in \Zbf, M \not= 0 \mod 5$. This provides a complement for $U_{3M}$, and hence a lift
for minimal models with $k=18p+1$ with the aforementioned restriction on $M$. The corresponding values of $k$ that occur in at least one of the
168  Gepner models are 1 and 19 only.

In the next case we have three  choices for the $A_6$ representation, so we get three equations of the same kind. For each of the
three choices we obtain an infinite series of liftable $k$-values, but only a few are relevant: 
If the
$A_6$ weight is $\frac37$ we get $k=26$ for $M=8$, if the weight is $\frac57$ we get $k=5, M=2$ and $k=54, M=16$ and for weight
$\frac67$ we find $k=12, M=4$. Finally, the third case has two possible values for the $A_4 \times A_2$ weight, namely $\frac{11}{15}$
and $\frac{14}{15}$. The former provides two relevant solutions, $k=13, M=2$ (the one found above corresponding to the standard model
$U(1)$) and $k=238, M=32$, plus the usual infinite sequence of irrelevant ones. The last value provides $k=58$ for $M=8$.

The different solutions are summarized in Table 1. The column ``Lift" indicates
which CFT is the complement of the denominator $U(1)$. All affine algebras are at level 1. The
last three columns indicate how many previously light ($h \leq 1$) states are lifted, how many
heavy ones are lowered to $h \leq 1$, and how many are light and remain light. These numbers depend
on the matching of the minimal model with its lift, which may not be unique. In general this
matching can be altered by any fusion rule automorphism of the CFT. Choosing a different
fusion rule automorphism leads to a different pairing of the primaries, and, since fusion
rule automorphisms only respect conformal weights modulo integers,  it may lead to different
counts for (un)lifted states. 
There does not exist a canonical choice for this automorphism,
and the numbers in the table only represent one of the choices. There is no need to consider
other choices as additional lifts, because all pure automorphisms of minimal models can be
obtained by means of simple currents. Hence they will be all be sampled if we scan the
simple current MIPFs. In the simplest case, $k=1$, the two Ramond states with $c=\frac{1}{24}$
are lifted to $c=\frac{25}{24}$, the two Neveu-Schwarz states with $c=\frac{1}{6}$ are lifted to
$c={7}{6}$, and the world sheet supercurrent is lowered to $c=\frac12$. In this case these numbers are unique. In all cases in the table we have explicitly checked equality of the complete matrices $S$ and $T$. 

\begin{table}
\begin{center}
\vskip .7truecm
\begin{tabular}{|c||c|c|c|c|}
\hline
\hline
$k$ & Lift & Lifted  & Lowered & Unchanged \\
\hline
1 & $E_6 \times A_1$ & 4 &  1 & 4 \\
2 & $A_7$ &  7  &  1  &  12  \\
3 & $[D_6 \times U_{10}]_{\rm ext}$ & 10  &  3  &  22 \\
4 & $D_5 \times A_2$ & 21  &  4  &  23 \\
5 & $A_6 \times A_1$ & 32  &  8   &  29  \\
5 & $ [E_6 \times U_{42}]_{\rm ext}$ & 24 & 11 & 37 \\
6 & $[ A_6 \times U_{112}]_{\rm ext} $& 33 &15 & 39 \\
8 & $A_4 \times A_3$ &  65 &  29    &  37 \\
9 & $ [A_6 \times U_{154}]_{\rm ext}$ & 76 & 41 & 39 \\
11 & $[E_6 \times U_{78}]_{\rm ext}$ & 104  &  61  &  39 \\
11 & $[D_6 \times U_{26}]_{\rm ext}$ &  98   &  60     &  45   \\
12 & $A_6 \times U_4$ & 125  &  66  &  39 \\
13 & $A_4 \times A_2 \times A_1 $ & 136  & 81   & 37  \\
14 & $[A_4 \times A_2 \times U_{480}]_{\rm ext} $ &  147  & 105   & 47  \\
14 & $ [A_6 \times U_{224}]_{\rm ext}$ & 153 &  95 &  41 \\
17 & $[E_6 \times U_{114}]_{\rm ext} $ &  202  & 105   & 37  \\
17 & $[A_4 \times A_2 \times U_{570}]_{\rm ext} $ &  198   &  133   &  41  \\
19 & $E_6 \times U_{14}$ &  228 &   119 &  42  \\
20 & $ [A_6 \times U_{308}]_{\rm ext}$ &  243 & 143  & 42  \\
23 & $[D_6 \times U_{50}]_{\rm ext}$ &  300  &  161   &  41  \\
26 & $A_6 \times U_8$ &  349 &  199  &  39  \\
30 & $ [A_6 \times U_{448}]_{\rm ext}$ & 417 & 235  & 46  \\
41 & $[E_6 \times U_{258}]_{\rm ext}$ & 610 &  297  &  44 \\
41 & $[A_6 \times U_{602}]_{\rm ext}$ & 606   &   325  &  48  \\
42 & $[A_6 \times U_{616}]_{\rm ext}$ &  627 &  337   & 46   \\
44 & $ [A_6 \times U_{644}]_{\rm ext}$ & 673 & 361  &42   \\
44 & $ [A_4 \times A_2 \times U_{1380}]_{\rm ext}$ & 659 &  465  & 56  \\
47 & $[E_6 \times U_{294}]_{\rm ext}$ & 728 &  367  &  46 \\
54 & $A_6 \times U_{16}$ &  857 &  455  &  51 \\
58 & $A_4 \times A_2 \times U_8$  & 923 & 611 & 56 \\
86 & $ [A_6 \times U_{1232}]_{\rm ext}$ & 1501  &  741 &  52 \\
89 & $[E_6 \times U_{546}]_{\rm ext}$ & 1556 &  705  &  49 \\
238 & $A_4 \times A_2 \times  U_{32}$ &  4959 &  2729  &  73  \\
1,1 & $A_2 \times A_1 \times A_2 \times A_1$ & 16 & 1 & 14 \\
\hline
\end{tabular}
\vskip .7truecm
\caption{List of all lifts of N=2 minimal models described in this paper.}
\end{center}
\end{table}

The embeddings discussed so far correspond to the entries in the table without an ``ext" suffix. The ones with such a  
suffix are obtained by applying an appropriate integral spin extension to the  tensor product.  Some of these cases were found by
observing that some $U_N$ factors are present  in column 2 for which no complement was found so far.
For example, it follows from the table that $E_8$ can be written as an extension of
$U_{480} \times U_{32} \times A_4 \times A_2$. This can be used directly  to lift the $k=238$ model, which has a $U_{480}$
denominator factor.  It cannot be used directly to lift the $k=14$ model ($U_{32}$). To construct the lift of the $k=14$ minimal model we merely have to extend $U_{480} \times A_{4} \times A_{1}$ with
an integer spin order 15 current to obtain a complement of $U_{32}$. In a similar way one can obtain a lift of the $k=6$
model and a second(!) distinct lift of the $k=5$  model\rlap.\footnote{The complement of a CFT is not necessarily unique, even
if the central charge is fixed.  There are many examples of this in the list of $c=24$ meromorphic CFT's} At first there also seems to exist an alternative lift of $k=2$,
 $[A_4 \times A_2 \times U_{120}]_{\rm ext} $, but this is  equivalent to $A_7$. 
 
 The extended solutions described above were obtained by interchanging the r\^oles of two $U$ factors in a product of the
 form $U_N \times U_M \times X_6$, where $X_6$ is a simply laced rank 6 algebra, and $U_N$ is the unitary factor that corresponds
 to a minimal $N=2$ factor that appears in a $c=9$ tensor product. 
 In the unextended cases in the table, $M$ is always smaller than $N$
 by a factor $y$, and $y$ is the number of primaries (and simple currents) of $X_6$, so that $yM=N$. 
 This means that if we interchange $N$ and $M$, we must extend $U_N  \times X_6$ by a simple current of order $y$ so that
 the total number of simple currents of $[U_M \times X_6]_{\rm ext}$ equals $\frac{Ny}{y^2}=M$. However, if we interchange $N$ and $M$
 it is $U_M$, and not $U_N$, that must correspond to a minimal model in  a $c=9$ tensor product. Hence one may expect additional
 solutions, and indeed there are. They correspond to the other lifts with $X_6=A_6, A_4\times A_2$ or $E_6$ listed in the table.
 
 But now it is clear that there are still more possibilities. We have explained above that if $X_6$ is an unextended tensor product,
 it should have an odd number of primaries. But if we allow extensions, that restriction is not necessary anymore, because
 the extension may reduce the number of primaries from even to odd. So now we may also consider $D_6$ or $D_4\times A_2$. Consider
 for example the combinations $U_{2M}\times U_{2M} \times D_6$. It is easy to check that they can be extended to $E_8$ for $M=5,13$ and $25$, and that this yields lifts for $k=3, 11$ and 23 (as in the previous cases, there is an infinite sequence of solutions, but the higher
 ones do not occur in $c=9$ tensor products).  It is likely that many more lifts can be constructed. The most general approach would be
 to start with eight copies of $U_{2M}$  with different values of $M$, whose product is a square, and look for simple current extensions that
 reduce it to one primary field. This would of course include all cases found above, with the simple algebras appearing as spin-1
 extensions. Unfortunately it is not at all clear what is the maximal value of $M$ that needs to be considered (if we limit ourselves
 to the relevant minimal models). The examples above contain factors as large as $U_{1380}$.  
 This implies that an  exhaustive search could turn out to be rather challenging. 
 
 But there are {\it still} more possibilities. In some cases one can lift two minimal model factors simultaneously by embedding both $U(1)$'s in $E_8$. It is important (as it is in the examples above) that this embedding be faithful: it must be possible to factorize
the $E_8$ partition function in terms of the full set of characters of the $U(1)$ factors, with each character combination
appearing. For example the aforementioned  tensor product  $U_{480} \times U_{32} \times A_4 \times A_2$
cannot be used
to get a complement for  $U_{480}\times U_{32}$: there is simply not enough information in $A_4 \times A_2$. There is
however at least one example of a double lift, namely $A_2 \times A_1 \times  A_2 \times A_1$, which is a complement
of $(U_6)^2$, and can be used to lift two $k=1$ factors. Many more may exist. There may also exist triple and 
quadruple lifts.

 Since multiple lifts may exist for the same
 minimal model, one may even ask if the total number of lifts per minimal model is finite! We do not have a proof that it is, but it
 is easy to see that the number of distinct massless spectra that can be obtained is finite. This can be seen as follows.
 There is only a finite number of regular
 Gepner models, and each of them has a finite number of primaries. For each primary, the only relevant data are its conformal weight
 and its ground state dimension. The value of the conformal weight $h$ is relevant for massless spectra only if $h\leq 1$, and rationality
 then guarantees that there is just a finite number of possible values. The dimensions
 are limited, because they must be contained in some $E_8$ representation.
 The maximal $E_8$ representation that can occur is determined by the excitation level within the $E_8$ character: a given ground state
 $i$ of the complement $X_7$ of $U_M$, with conformal weight $h_i < 1$, combines with a $U_M$ primary $q$ to give an $E_8$ excited state. The conformal weight $h_q$ of $q$ is limited by the value of $M$, and hence the value of $h_q+h_i$ is limited as well.

The total number of lifted Gepner combinations that can be obtained with the single lifts in the table is 435. 
We have examined about 50 of them until now, and found
three-family  models in half of them. So far no three family models were found using double lifts.

We distinguish spectra on the basis of criteria similar to those used in earlier work on standard model spectra in orientifold models \cite{Dijkstra:2004cc}. We ignore the details of the ``hidden" gauge group (which is defined to be anything outside $SO(10)$, and is not necessarily truly hidden), and
only record the total dimensions. We only consider spectra where the chiral states can be combined into standard model families.
All others are counted as well, but not distinguished. 
Spectra are considered distinct if the standard model is embedded in a different subgroup of $SO(10)$,
if $B-L$ is or is not anomalous, if the number of mirrors of the charged quarks and leptons is different, and if the number of standard
model singlets is different. Furthermore, if $SO(10)$ is broken to a subgroup, there will always be fractionally charged states
in the spectrum ({\it i.e.} states which can form fractionally charged color singlets). 
By the foregoing assumption, we will not consider any spectrum in detail if those fractionally charged states are
chiral. If they are non-chiral and massless, we only distinguish spectra on the basis of their total dimension (we have not encountered any
examples where all these states are massive; further details will be presented in \cite{GS}). 

In most cases we have seen, the number of distinct three family spectra (if any) is a few tens or hundreds. 
Most of these have just been scanned superficially to establish the existence of three family models, and hence we only have a rough
estimate of the total number that they may contain. So far, 
the combination
$(\hat6,6,6,6)$ produced $ > 286$ distinct three-family models, $(\hat2,2,2,2,2,2)$ yields  $ > 650$  and $(\hat1,1,1,1,1,1,1,1,1)$  $ > 162$. 
The latter ones are of special interest,  since they can be realized completely in terms of free bosons: both the $k=1$ model and
its lift can be realized in that way. Hence the three family models in this class are realizable using the ``covariant lattice construction"
of \cite{LLS}.  
We also studied the $(5,5,5,12)$ combination to compare the two distinct lifts of the $k=5$ model and the lift
of the $k=12$ model. They do indeed yield different results. The combinations $(\hat 5,5,5,12)$ and $(5,5,5,\widehat{12})$
both have family number quantized in units  of 1, and at most 8 families, but the distributions and the spectra are quite different; 
the $(\tilde5,5,5,12)$  combination ($\tilde5$ denotes the second lift) has family number quantized in multiples of 3 with a maximum of 6. 
All three combinations yield three family models, about a hundred in total. 
But a few combinations give orders of magnitude more. The combination ($\hat3,3,3,3,3)$ has more that 3500 distinct three-family spectra.
The largest set so far comes from the combinations $(3,\hat8,8,8)$, which has produced more that 5700 distinct 3-family spectra. 
The combination $(\hat3,8,8,8)$ produced more than 4000 so far. 

We conclude with a few more details about just one of the 450 lifted Gepner models that we have available now.
The  $(3,\hat8,8,8)$ combination has spectra with any number of families up to 20, and all even numbers up to 32. A plot of the family distribution is shown in figure 1. 

\begin{figure}
\begin{center}
\includegraphics[width=17cm]{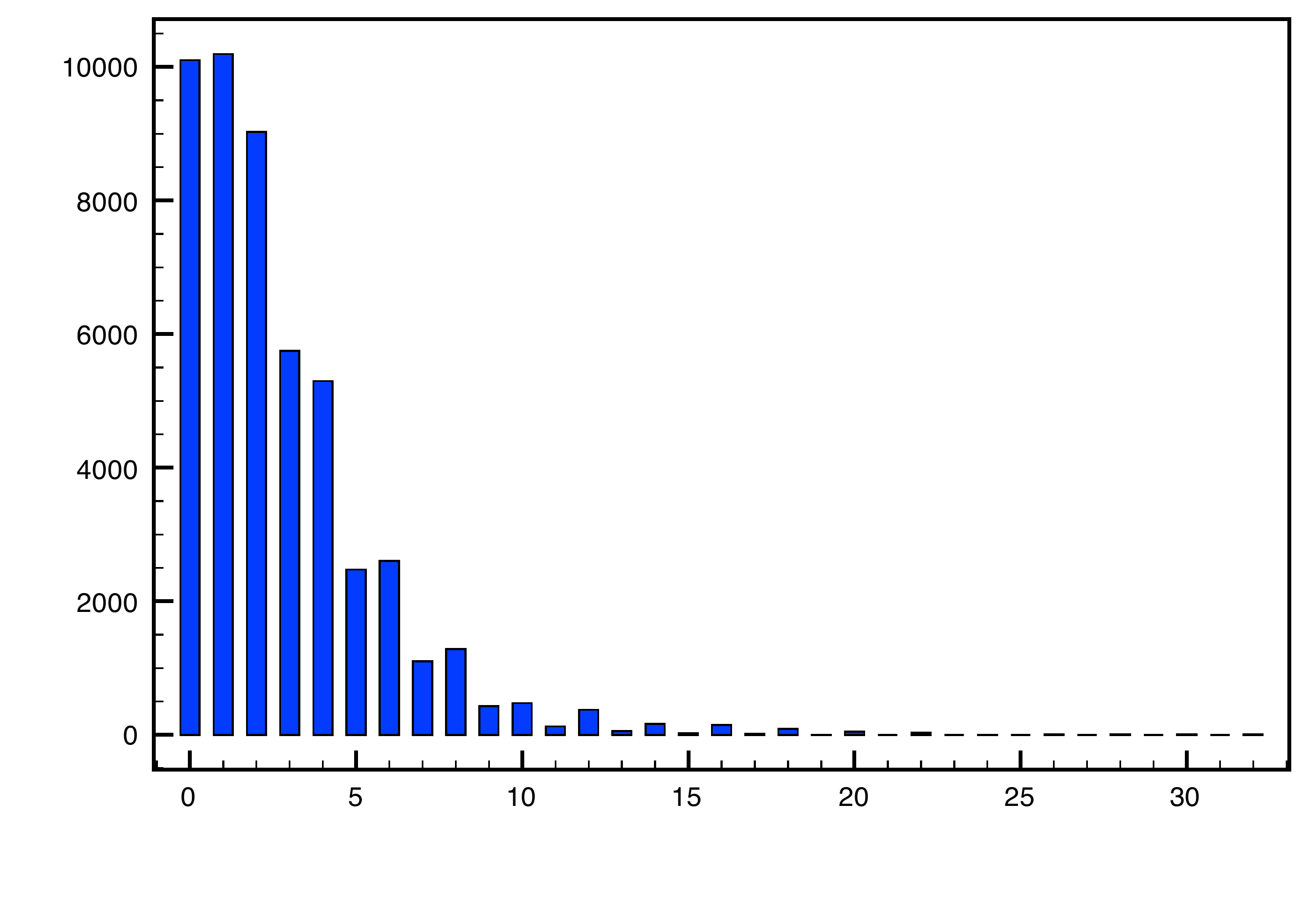}
\vskip -1.cm
\caption{Distribution of the number of families for the $(3,\hat8,8,8)$ tensor product. The total number of spectra is about 50000.}
\end{center}
\end{figure}

This is probably the first time that a really fine-grained family distribution can be plotted for heterotic strings (statistical
results focusing on gauge groups, but without requiring standard model families were published in \cite{Dienes:2006ut}).
Note that this is
based on just one of the hundreds of tensor products we can consider. It is noteworthy that the plot shows a sharp, approximately
exponential decrease with increasing numbers of families, just as the analogous plots for orientifolds. 
However, the orientifold family plot falls off a lot faster than the heterotic one.
Note also a
slight preference for even numbers of families, which however is not nearly as pronounced as it was for the Gepner orientifolds \cite{Dijkstra:2004cc} or the $\Zbf_2 \times \Zbf_2$ orientifolds 
\cite{Gmeiner:2005vz}. A consequence of both these effects is that the fraction of three family models is 
much larger in lifted Gepner models 
than it is in orientifold models (although the latter are far more numerous). 
The exponential behavior for other combinations looks similar, but
it remains to be seen if this feature continues to hold in the full set of lifted Gepner models. 

Most of the 5700 three family spectra have a gauge group $SO(10)$ or $SU(5)$. The number of mirrors can range from 0 to 30,
and there are from 200 to 500 singlets. It is unlikely that all of these 5700 models are unrelated. Most likely what we are seeing here is a discrete scan over a moduli space, and the large number of discrete spectra is
due to different numbers of mirror fermions, singlets and other non-chiral particles being lifted at
special points in that moduli space. The same phenomenon has been observed in RCFT orientifolds,
although in that case the interpretation is more obvious: it is a consequence of changing boundary states, which in a geometric interpretation corresponds to moving branes. In the present situation it remains to be elucidated what the moduli space of these models is, assuming it exists.

The spectrum with the smallest number of mirrors is an $SO(10)$ model with a 
spectrum that is just  $ 3 \times (16) + 1 \times (10)$. It comes in two varieties, with 460 or 467 singlets. Apart from the large
number of singlets, this is {\bf \underline{exactly}} the matter content of the minimal $SO(10)$ Susy GUT, since the $(10)$ can be identified as the multiplet containing
the standard model Higgs. It should be noted, however, that the massless spectrum contains no Higgs to break $SO(10)$ to the
standard model. This is always the case in heterotic GUT models based on level-1 affine algebras, and can be solved by
breaking $SO(10)$ directly at the string scale, but only at the expense of fractionally charged particles. 

We do indeed get examples where $SO(10)$ is broken, including cases where it is broken down completely to
$SU(3)\times SU(2) \times U(1)$ and where $B-L$ is anomalous (of course there are the usual additional gauge group factors that do
not originate from $SO(10)$). The two simplest cases  have no mirrors for the usual $Q$, $U^c$ and $E^c$ standard model representations,
3 $D^c$ and 3 $L$ mirrors, or 2 $D^c$ and 9 $L$ mirrors. We will describe the first one in a little more detail. The additional
affine Lie algebra remaining from the broken $E_8$ is $A_{1,8} \times U_4 \times A_{3,1} \times A_{4,1}$, {\it i.e.} the extra gauge group is
$SU(2)\times SO(2)\times SU(4) \times SU(5)$. All massless matter may be charged under this gauge group. For the quarks and
leptons we find that the 3 $U^c$'s owe their multiplicity to being in the spin 1 triplet representation of $SU(2)_8$ (the subscript indicates the level).
 The same is true for the three $E^c$'s and three of the
six $D^c$'s; the three others are singlets. All other quarks and leptons 
are singlets as well. In particular all weak lepton doublets and their mirrors, some of which might play the r\^ole of standard
model Higgs bosons, are singlets of $SU(2)_8$. 
Therefore in this particular example all up quark and charged lepton Yukawa couplings are forbidden as long as the $SU(2)_8$ is unbroken. 
However, this is just
a fairly randomly chosen example, and all these features
will strongly depend on the case one considers. In comparison to the state of the art of heterotic model building based on Gepner 
models until now, it is already miraculous that we can start discussing Higgs couplings at all. 

In this example, the anomalous $B-L$ charge is not universal for quarks and leptons in different families. In fact $B-L$ is a misnomer here, and
refers simply to the additional $U(1)$ in $SO(10)$. It turns out that the quarks and leptons that are in the triplet
representation of $SU(2)_8$
are uncharged under this $U(1)$, whereas all other quarks and leptons have charges $-2,0$ and 2.

One of the problems where weight lifting might help is the moduli problem. Unfortunately it is not easy to study that in an exact
RCFT point and without a known geometric interpretation. But we can examine the singlets in the massless spectrum. 
Above we have seen that in a rational scan of models we can remove almost all mirror fermions. Can the same be done with
the singlets? So far we have found that the number of singlets does indeed take many different values over the 
rational ``moduli space", but it does not even get close to zero. 
The total number of standard model singlets in the example discussed above is 250. Of these, 38 are also singlets of $B-L$, and hence singlets
under all generators of the original $SO(10)$. This is substantially less than the number of $E_6$ singlets in the
corresponding regular Gepner model (3,8,8,8), namely 495. Furthermore the singlets in the lifted models are in non-trivial
representations of the additional gauge group. In this case, the decomposition of these 38 singlets under the additional 
gauge group is as follows
$$ (5,0,0,0)+(3,0,0,0) + 3 \times (0,v,0,0) + (3,0,0,5) + (0,0,6,0) + 3 \times (0,0,0,0)\ , $$
where we have indicated $SU(2)_8$,  $SU(4)$ and $SU(5)$ representations by their
dimension and $v$ denotes the vector representation of $SO(2)$. There are just three absolute singlets, but one should keep
in mind that many moduli in standard Gepner models are charged under the $U(1)$'s corresponding to the minimal model factors. 
In the present situation such charges will be converted into representations of the extra gauge groups. 

The spectrum of this model contains the expected fractionally charged particles, but only with half-integer charge
(in general third-integral or sixth-integral is also possible).  We require this set of states to be non-chiral in $SU(3)\times SU(2)\times U(1)$,
since otherwise we would not even count the spectrum as an N-family model.
The total dimension of this set of states is 172, and remarkably they
are all in half-integral spin representations of $SU(2)_8$ including one with
spin $\frac72$. We have checked that all integrally charged massless particles, including all singlets,  
are in integral spin representations of $SU(2)_8$. 
Consequently in this model all fractionally charged particles can be confined into integrally charged particles by $SU(2)_8$!
Dynamically, it is quite a bit more complicated than that. Note that some of the quarks and leptons are triplets of this $SU(2)$, and hence
are affected if it becomes strong. Furthermore we have to check that the full spectrum, not just the massless one, has this feature. If it
is just a coincidental feature of the massless spectrum (which seems unlikely), this may already be good enough, but  there is a chance that it is a 
general feature that occurs in many other cases as well. This is certainly an issue we will investigate in the future, but it is far beyond
the scope of the present paper, namely explaining the construction and examining the possible existence of chiral spectra.

It is noteworthy that the aforementioned fractional charge confinement mechanism uses one of the gauge groups that emerged as a result of the lift, and hence would not work for normal Gepner models, even if they had three families. It is also noteworthy that this happened for the first
example we examined in detail.
With so many nice features emerging  spontaneously and
abundantly, one cannot avoid the feeling that after two decades in a barren valley of the heterotic RCFT landscape we
have finally reached the huge fertile plane we were once hoping for. The
heterotic RCFT landscape is now ready for the same kind of systematic explorations that have been done extensively in the case of orientifolds (see {\it e.g.} \cite{Dijkstra:2004cc}\cite{Gmeiner:2005vz}\cite{GatoRivera:2005qd}\cite{GatoRivera:2008zn}\cite{adks}\cite{Gmeiner:2008xq}), and
is showing some  similar features, such as the distribution of families. 

This construction raises numerous questions, such as: 
what, if any, is the geometric interpretation of these models? Are they related to other constructions, and how?
Is there a related Landau-Ginzburg description \cite{Greene:1988ut}?
What are their strong coupling duals? Is there an exact mirror symmetry? 
Is it possible to classify all the lifts? 
Are there any generic bad features that rule out this entire class
phenomenologically? What can be said in general about charge quantization and confinement? Is there a simple
rule for family number quantization?
How close can we get to the MSSM spectrum?  Without supersymmetry, how close can we get to the SM spectrum?

We hope to answer some of these questions in the future.

\vskip 2.truecm
\noindent
{\bf Acknowledgements:}
\vskip .2in
\noindent
One of us (A.N.S) wishes to thank Massimo Bianchi for expressing his interest in
asymmetric Gepner models, which was part of the inspiration for this work.  This work has been partially 
supported by funding of the Spanish Ministerio de Ciencia e Innovaci\'on, Research Project
FPA2008-02968, and by the Project CONSOLIDER-INGENIO 2010, Programme CPAN
(CSD2007-00042). The work of A.N.S. has been performed as part of the program
FP 57 of Dutch Foundation for Fundamental Research of Matter (FOM).

\vskip .5in

\bibliography{REFS}
\bibliographystyle{lennaert}

\end{document}